# Comparison of Tomographic Reconstruction Algorithms for Infrared Imaging Video Bolometer Diagnostic in Plasma Devices

Vinit Pandya[1], Santosh P. Pandya[2], Ansh Patel[3], Kumudni Tahiliani[2] and Kumar Ajay[2]

[1]*Sardar Vallabhbhai National Institute of Technology (SVNIT), Surat-395007, India.*
[2] *Institute for Plasma Research (IPR), Near Indira Bridge, Bhat, Gandhinagar– 382428, India.*
[3]*Max Planck Institute for Plasma Physics, Garching- 85748, Germany.*

***Abstract -*** Infrared Imaging Video Bolometer (IRVB) measures total radiation power loss from plasma in 2 dimensions through a pinhole camera geometry. Where a free-standing thin metal foil act as a broad band absorber from Soft X-Rays to IR radiation. This configuration produces line-integrated signals with poloidal and toroidal coverage that must be inverted to recover the plasma radiation emissivity distribution on a poloidal cross-section. This study compares the tomographic methods implemented to IRVB brightness data reconstruction, namely Minimum Fisher Information (MFI), Phillips-Tikhonov regularization (PTR), and Maximum-Likelihood Expectation-Maximization (MLEM). The comparison assessment is organized around several aspects of bolometer measurements, namely viewing geometry configuration, non-negativity, robustness to noise, sensitivity to prior assumptions, convergence speed, and peak preservation. The present work also details the IRVB forward modelling process, construction of synthetic phantoms, and a validation of these reconstruction methods based on typical expected emissivity profiles, namely symmetric Gaussian distribution at plasma center, symmetric hollow-radiation emissivity profile, asymmetric radiation profiles across the poloidal cross-section, and divertor-side radiation emission profiles. The outcome is to emphasize the practical tradeoffs among reconstruction accuracy, numerical stability, and suitability for real-time or offline usage of these reconstruction methods, particularly for the IRVB camera viewing system.

## I. Introduction

BOLOMETERS are currently primary diagnostic tools used to estimate the total radiated power in magnetically confined plasmas. There are a couple of types of bolometers utilized for the bolometric measurements on plasma devices, namely: Metal Foil Resistive Bolometers (MFRB) [1], [2], [3], Absolute eXtended UltraViolet (AXUV) diodes-based bolometers [4], [5], Infrared Imaging Video Bolometers (IRVB) [6], [7], [8], [9], Single channel metal foil IR-bolometers [10], [11], Fiber Optic Bolometers (FOB) [12], [13].

From these different bolometric measurement techniques, the Infrared Imaging Video Bolometer (IRVB) has emerged as a potential imaging diagnostic technique for measuring the total radiated power loss from magnetically confined plasmas. It can provide spatially and temporally resolved 2D profiles of plasma radiation from the plasma devices. The IRVB technique has been deployed on several plasma devices worldwide and has also been proposed for the ITER tokamak [8], [14]. This technique provides several important advantages, including immunity to electromagnetic noise from optical signal transmission, a wide spectral response band, radiation hardness for nuclear fusion devices [14], a wide dynamic range for power-loss measurements, and improved tomographic reconstruction of emissivity profiles with fewer IRVB systems. The IRVB diagnostic systems have been designed and developed at the Institute for Plasma Research (IPR) for the ADITYA, SST-1, and ADITYA-Upgrade tokamaks [7], [15], [16].

In an InfraRed Imaging Video Bolometer (IRVB), a thin absorbing metal foil is exposed to plasma radiation through a pinhole camera geometry, and each bolometer pixel on the foil corresponds to a distinct line of sight (LoS) passing throughout the plasma and it samples a finite volume of plasma. Upon incidence of radiation power from the sampled volume by the bolometer pixel foil, the foil gets heated up. An infrared (IR) camera placed outside the vacuum-vessel registers the temperature evolution during the plasma discharge. The infrared camera, therefore, records a set of line-integrated measurements which, by use of a suitable tomographic method can yield approximate local emissivity map. Recovering the emissivity distribution in the poloidal plane is an ill-posed inverse problem and several methods have been tested so far for AXUV [17] and metal foil bolometer [18] brightness measurements. In the case of IRVB, these inversion methods were applied for tomographic inversion of 2D radiation brightness images namely Minimum Fisher Information (MFI),

†This work was performed when the author was affiliated to Sardar Vallabhbhai National Institute of Technology, Surat; Corresponding author email ID: vdpandya2002@gmail.com

Phillips-Tikhonov regularization (PTR) applied on HL-2A [19], [20], [21], KSTAR [22], [23], [24], [25], JT-60U and JT-60SA , machine learning method on KSTAR [26]. However, in all these investigations, relatively large array of bolometer pixels has been considered. In present study, these methods are applied to the IRVB system deployed on the ADITYA tokamak where relatively a smaller number of bolometer pixel array is available (9 x 9 bolometer 2D pixel array).

In order to investigate applicability of these methods, the present study focuses on three primary methods for comparison: Minimum Fisher Information (MFI), Phillips-Tikhonov regularization (PTR), and Maximum Likelihood Expectation Maximization (MLEM) [27], [28]. For IRVB, the MLEM is considered and applied for the first time in this work. These methods are well suited for reconstruction of IRVB data because they can incorporate the finite detector geometry, enforce non-negativity, and compensate for sparse or limited-angle view coverage. Other inversion methods such as natural basis functions, SVD, maximum entropy method, etc. are not considered as they are relatively less effective than the above methods [29], [30]. Investigation on machine learning methods for small- to medium-sized plasma devices will be scope for future work.

In this present paper, the reconstruction problem is formulated on a discrete grid with a forward model emissivity added with noise factors for real-time scenario suitability in medium-sized tokamaks. A comparative study across major algebraic tomographic algorithms is conducted for the first time. The IRVB geometry differs from conventional AXUV and Metal Foil Resistive bolometer arrays in an important way. The response matrix is typically denser and three-dimensional, because the camera observes a spatially distributed foil rather than a just a vertical coverage by a number of discrete chords. This provides richer spatial sampling, but it also makes the detector response more sensitive to calibration, finite pinhole size, and the mapping from camera pixels to plasma coordinates. The paper, therefore, treats geometry construction, phantom design, and algorithmic reconstruction as a single workflow.

The present paper is organized as follows, Section II covers IRVB geometry, formulation of different phantom for emissivity reconstruction test is covered in section III, results and error analysis for the reconstruction algorithms is made in section V followed by a study of number of bolometer channel and reconstruction grid size sensitivity of reconstruction algorithms across phantoms is reported in section VI.A and VI.B respectively.

## II. IRVB Geometry

In the present tomography inversion study, viewing geometry and typical parameters of the IRVB diagnostic system deployed on the ADITYA-Upgrade (ADITYA-U) tokamak [31], [32] have been considered. The typical plasma parameters of the ADITYA-U tokamak are summarized in Table-I.

TABLE - I
TYPICAL PARAMETERS OF THE ADITYA-U TOKAMAK

| Parameter | Values |
|---|---|
| Major radius ($R_0$) | $0.75\ m$ |
| Minor radius ($a_0$) (Limiter Plasma Configuration) | $0.25\ m$ |
| Plasma current ($I_p$) | $100 - 200\ kA$ |
| Line-averaged density $< n_e >$ | $1 - 3 \times 10^{19}\ m^{-3}$ |
| Electron temperature $T_e$ | $up\ to\ 500\ eV$ |

The IRVB system in ADITYA-U [7] employs a graphite-coated (IR-camera side) free-standing platinum foil (63 × 63 mm², 2.5 μm thick) as the radiation absorber, providing a broad and nearly flat spectral response to the emitted photons from plasma in energy range ~2 eV to ~5 keV. This range includes visible, ultra-violate (UV), and soft x-rays photons. Because of the finite aperture size, charge-exchange (CX) neutrals escaping from the plasma can also contribute to the signal. To form a pinhole camera geometry, the foil is kept 35 mm behind the pinhole flange, having a square aperture of 7 × 7 mm² and the pinhole aperture is 60 mm away from the LCFS (Last Closed Flux Surface) defined by the limiter. This pinhole-camera geometry forms a 9 × 9 bolometer pixel array (81 channels), giving ~6 cm spatial resolution at the plasma mid-plane. The foil temperature distribution is recorded using a 320 × 240 MWIR (Medium Wave Infrared) focal plane array camera operating at ~128 Hz (~8 ms temporal resolution) with ~25 mK noise equivalent temperature difference (NETD) of IR-camera, corresponding to ~200 µW/cm² noise equivalent power density (NEPD). The incident radiated power on the foil is obtained by solving the two-dimensional heat diffusion equation. The IRVB results show good agreement with AXUV bolometers and synthetic diagnostic modelling and have been used to study radiation enhancement during impurity injection and disruption mitigation experiments. Figure 1 shows the schematic layout of the IRVB components and their configuration.

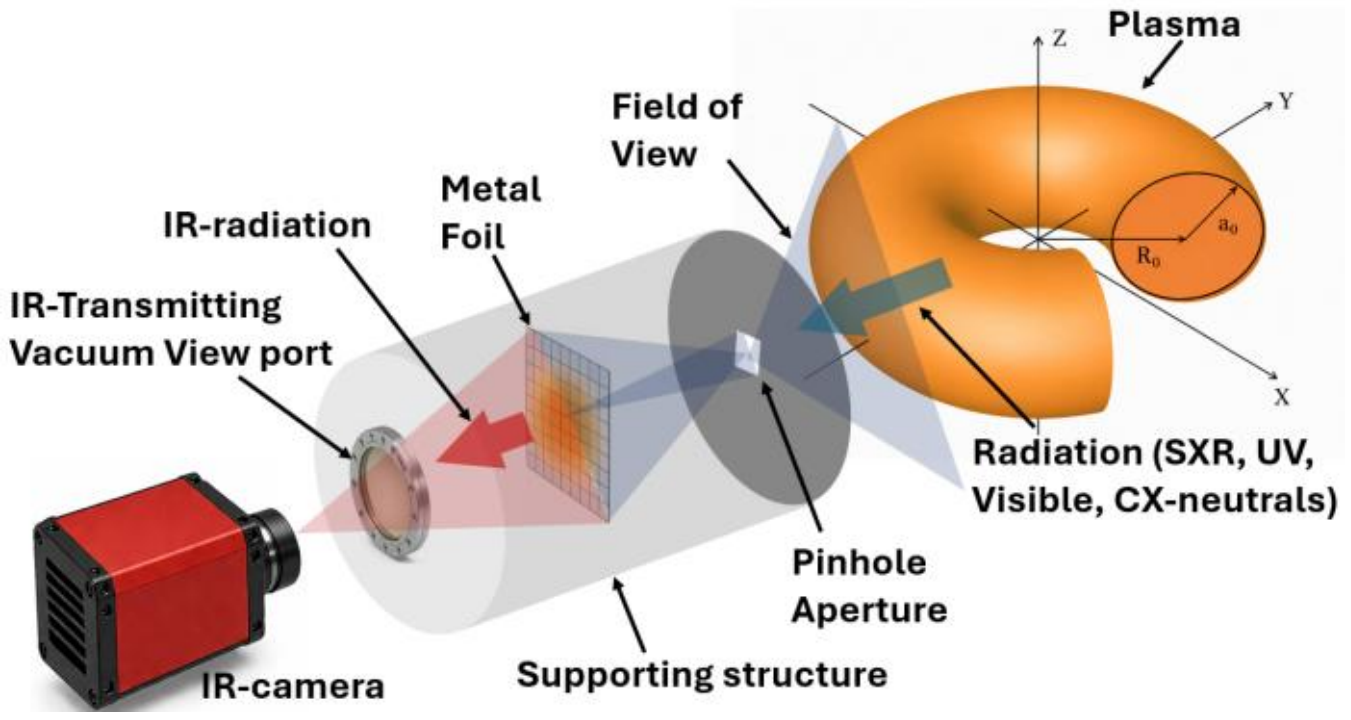

Figure 1. IRVB camera setup

It should be noted that the location of the pinhole relative to the foil center is an important parameter that facilitates the viewing geometry of the bolometer line of sight (LOS) into the plasma. The pinhole-aperture is located 59.5 mm away from the foil center. This forms a tangential viewing field of view (FOV) of the system. Thus, one column of the 9 bolometer pixels covers a complete poloidal cross-section. In contrast, the remaining bolometer pixel provides a tangential view of the plasma volume, as shown in Figure 2. Recently, the IRVB system on the ADITYA-U tokamak has been upgraded with a new high-performance IR-camera having 640 x 512 pixel format, 520 Hz frame rate, and a larger foil size as summarized in Table-2. The new upgraded IRVB system can provide a 13 (horizontal) x 15 (vertical) bolometer pixel array (195 channels).

For IRVB systems, the geometry matrix is usually constructed from ray tracing or detector point-response integration, not simply from idealized straight lines. In this study, the geometry model is obtained by discretizing the plasma cross-section into 3D voxels in $(R, \theta, \phi)$ where $\phi$ is the toroidal angle. These toroidal co-ordinates can be transformed into cartesian co-ordinates $(x, y, z)$ by,

$$x = (R_0 + r cos\theta) \times cos\phi \quad (1)$$
$$y = (R_0 + r\cos\ \theta) \times \sin\phi \quad (2)$$
$$z = r\sin\ \theta \quad (3)$$

where, $R_0$ is the major radius, $r$ is the minor radius, $\theta$ is the poloidal angle, and $\phi$ is the toroidal angle. Computing the line-integrated contribution of each cell to every camera pixel needs the contribution of lengths of each LOS in each of these voxels. Since the voxels are 3-dimensional elements, the regular procedure of finding the lengths of LOS by boundary discretization can become computationally costly. Hence, a faster Siddon algorithm [4] is used where the 3D volume of each voxel can be considered as an intersection of three sets of orthogonal, equally spaced parallel planes. Instead of testing every voxel in a volume, it determines where a ray intersects these planes to find the specific voxels it passes through and the length of the ray segment within each of them can be calculated. Figure 2Figure 2 A chord of bolometer pixel in 4th row and 8th column LOS passing through the voxels in **(x,y,z)** through a toroidal segment. shows a representative case of a chord of bolometer pixel in 4th row and 8th column LOS passing through the voxels in (x,y,z) through a toroidal segment.

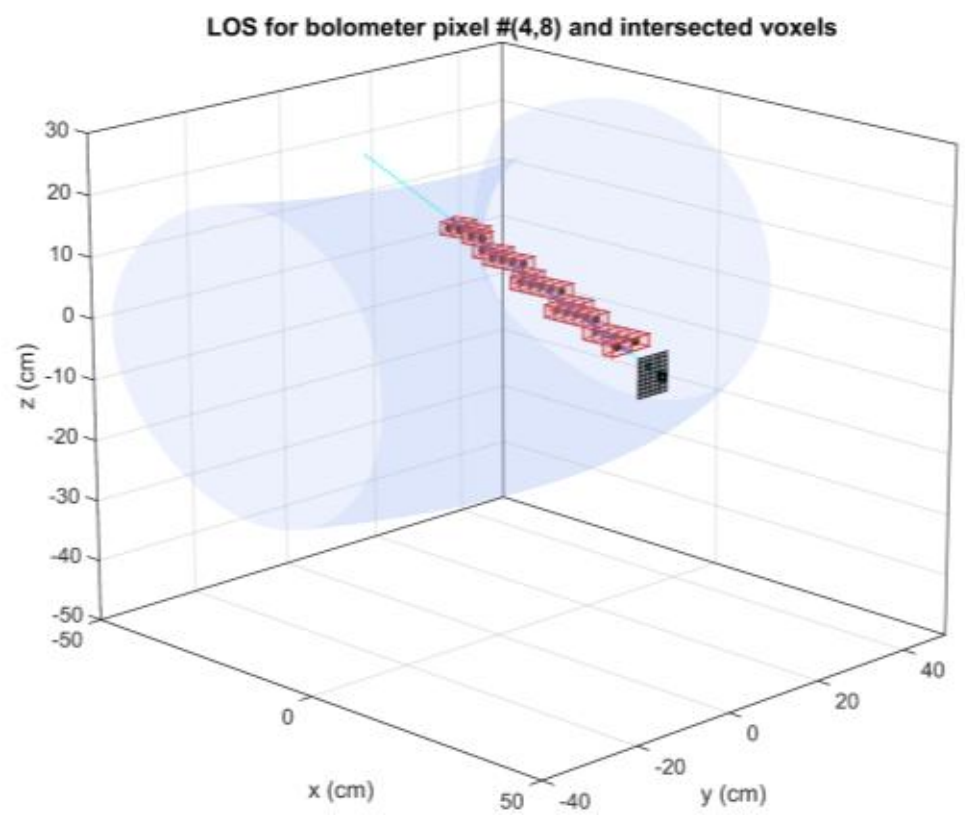

Figure 2 A chord of bolometer pixel in 4th row and 8th column LOS passing through the voxels in **(x,y,z)** through a toroidal segment.

The lengths of each LOS constitute the Geometry Matrix $L_{ij}$ where $i$ denotes the total number of available measurements $(\boldsymbol{f_x} \times \boldsymbol{f_y})$, in this case, the line-integrated emission, and $j$ denotes the lexicographically arranged voxels in $(R, \theta, \phi)$ directions. Figure 3 shows the geometry of a $9 \times 9$ IRVB LOS passing through voxel grids (represent as centers). This grid of voxels is constructed by $25 \times 25$ pixels in R-Z plane and by toroidally extending the plane in 110 wedges[1] from $\phi = 0$ to $\phi = 2\pi/3$ where the longest LOS extends inside the torus. The choice of number of pixels in the R-Z plane plays an important role in the emissivity reconstruction accuracy which is shown in the later section VI.

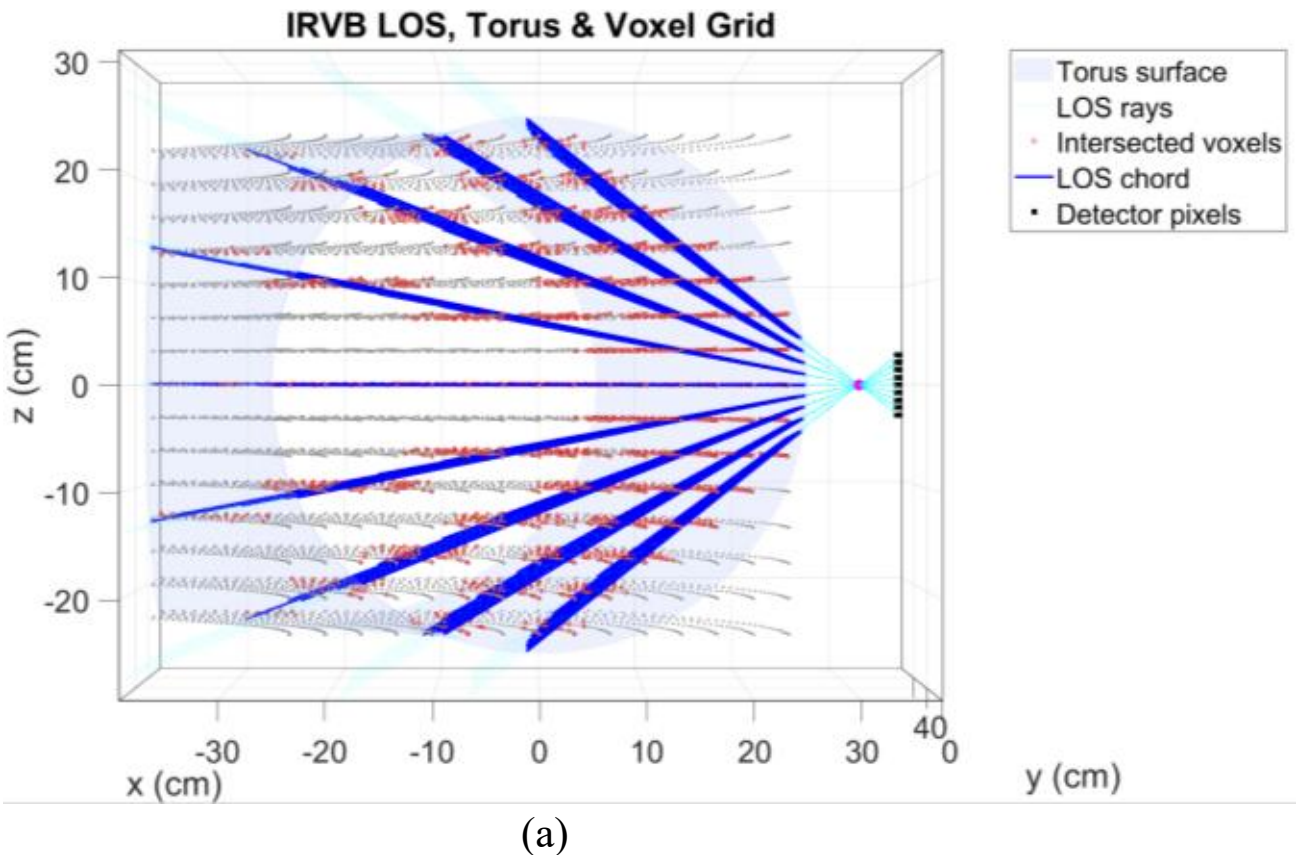

(a)

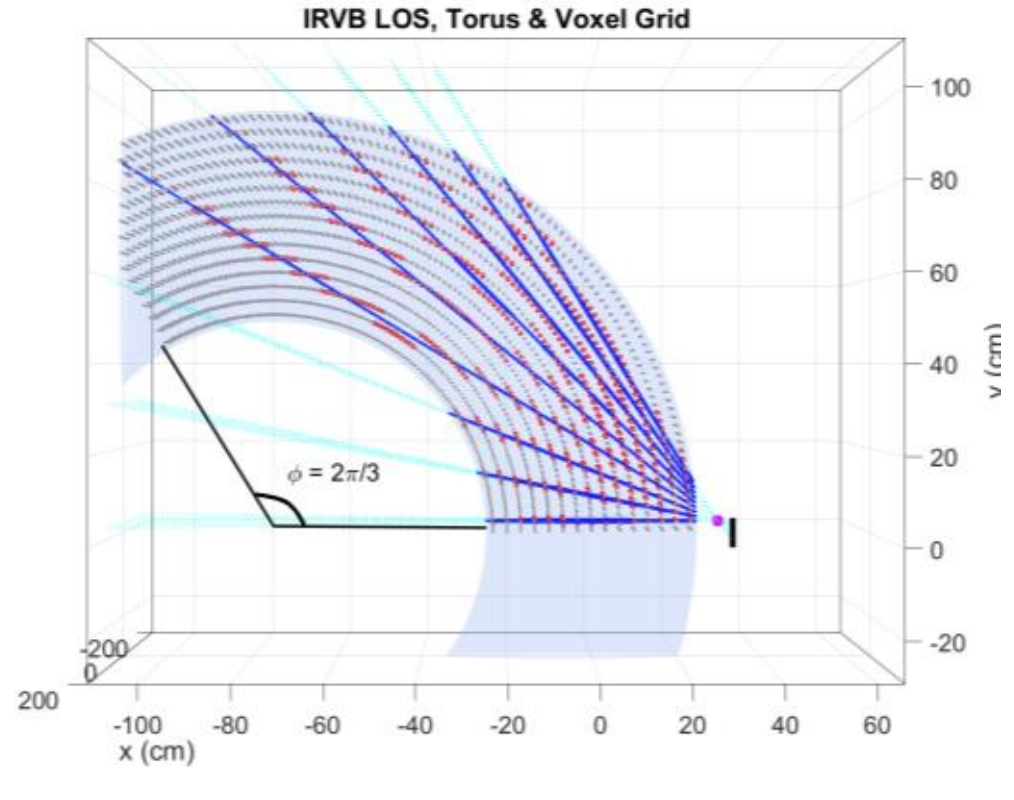

(b)

Figure 3. (a) Top and (b) Poloidal viewing geometry of IRVB LOS.

The geometry matrix $L_{ij}$ is a highly sparse matrix and to reduce the sparsity we can convert the 3D voxels into 2D projection plane. This can be achieved by summing up the 110 toroidal planes from $\phi = 0$ to $\phi = \frac{2\pi}{3}$ into one single poloidal plane, with the assumption that the radiation from plasma is toroidally symmetric. So, we create a new projection matrix as,

$$L_{proj} = \sum_{\phi=0}^{\phi=\frac{2\pi}{3}} L_{R\times Z\times\phi} \quad (4)$$

[1] It can be noted here that the number of toroidal wedges here contributes to the finer sampling of the transverse LOS so a high number is desirable.

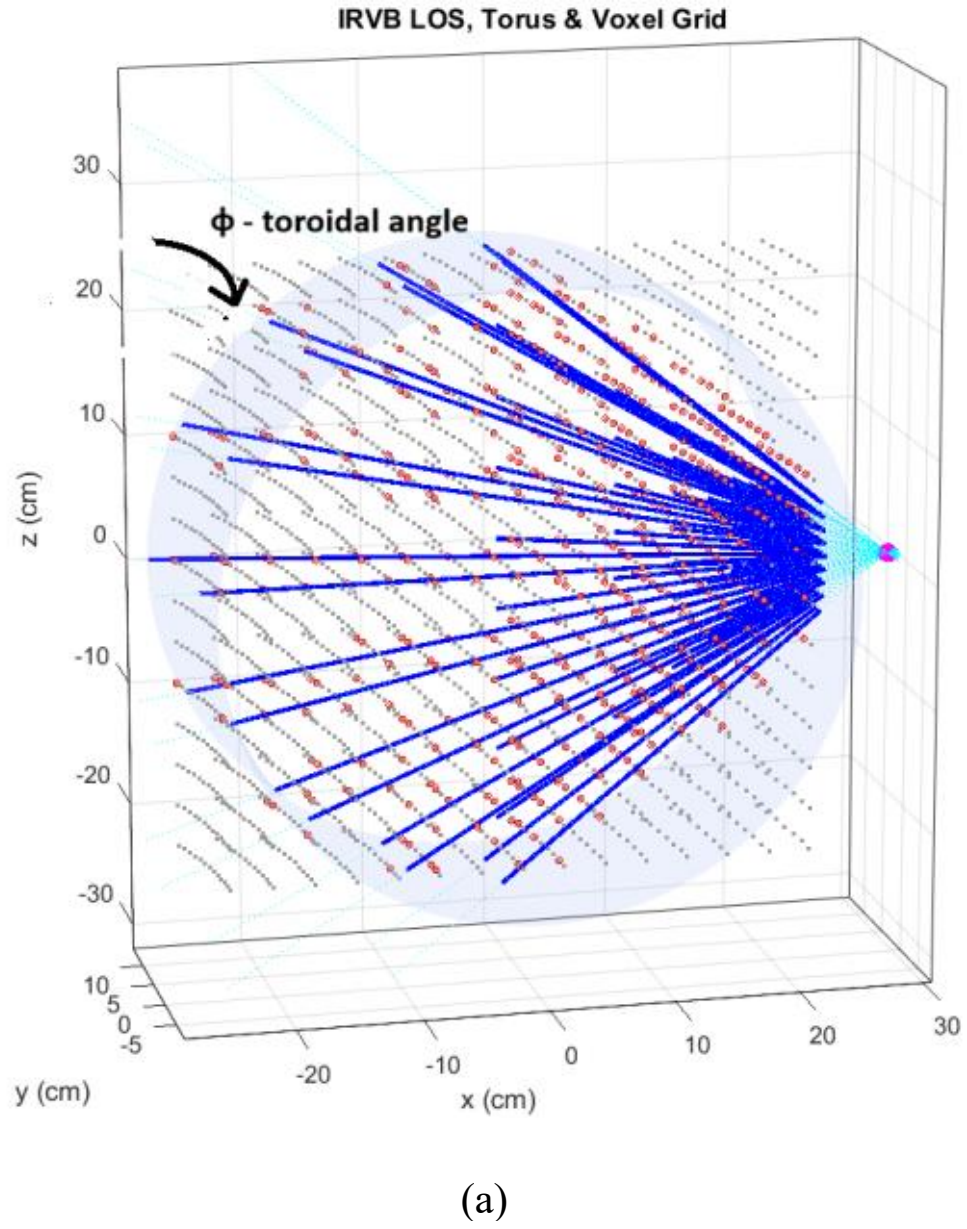

(a)

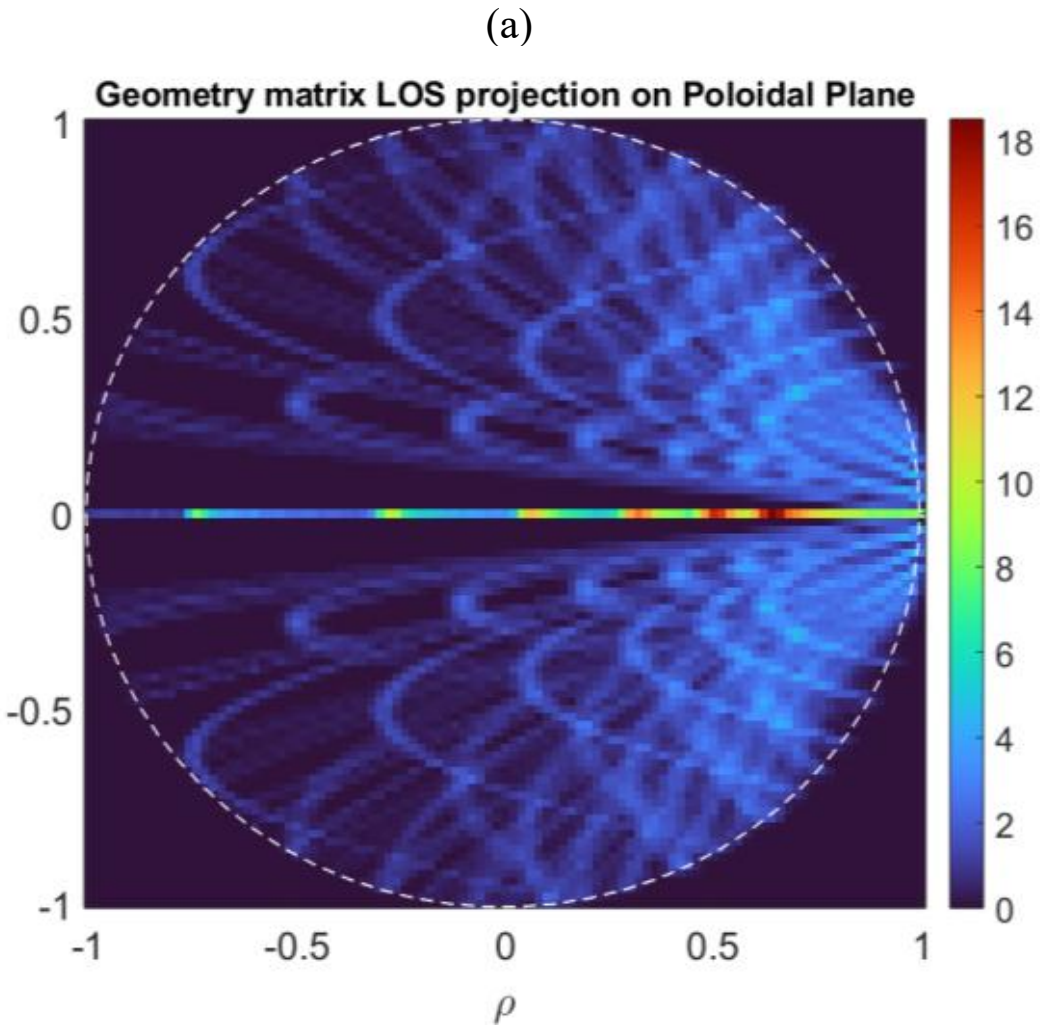

(b)

Figure 4. (a) The toroidal angle in the IRVB LOS viewing geometry and (b) The projection matrix $L_{proj}$ as shown in a poloidal plane when summed up.

## III. Forward Modelling

In matrix form, the relation between emissivity and measurements is

$$\boldsymbol{f} \;=\; \boldsymbol{L}_{\mathbf{proj}} \times \boldsymbol{E} \;+ \boldsymbol{\eta} \tag{5}$$

Where, the projection matrix $\boldsymbol{L}_{\mathbf{proj}}$ embeds the projection of contribution of LOS from each toroidal pixels summed up in a single poloidal plane, and $\boldsymbol{E}$ represents the 2D emissivity in the R-Z plane and $\boldsymbol{f} \in \mathbb{R}^{\mathrm{N}}$ is the measured brightness vector. The finite size of the foil, the aperture, and the thermal diffusion in the foil can all broaden the effective detector response. To consider all the errors in the measurement, an error function $\eta$ is introduced. Consequently, the inversion must handle both geometric and physical smoothing and minimize $\eta$. Five representative phantoms are constructed to span the radiative topologies that may typically be encountered in limiter and divertor tokamak operation.

### A. Core Gaussian Profile

The core Gaussian phantom represents centrally peaked emission characteristic of bremsstrahlung-dominated or medium-Z impurity-seeded plasmas:

$$\epsilon_1(R,Z) \;=\; \epsilon_0\, exp\left[-\frac{(R-R_0)^2+(Z-Z_0)^2}{2\sigma^2}\right] \tag{6}$$

with peak emissivity $\epsilon_0 = 1.0$ at the magnetic axis $(R_0, Z_0) \;=\; (0, 0)$ cm and width $\sigma \;=\; 8$ cm. This rotationally symmetric profile serves as the baseline case.

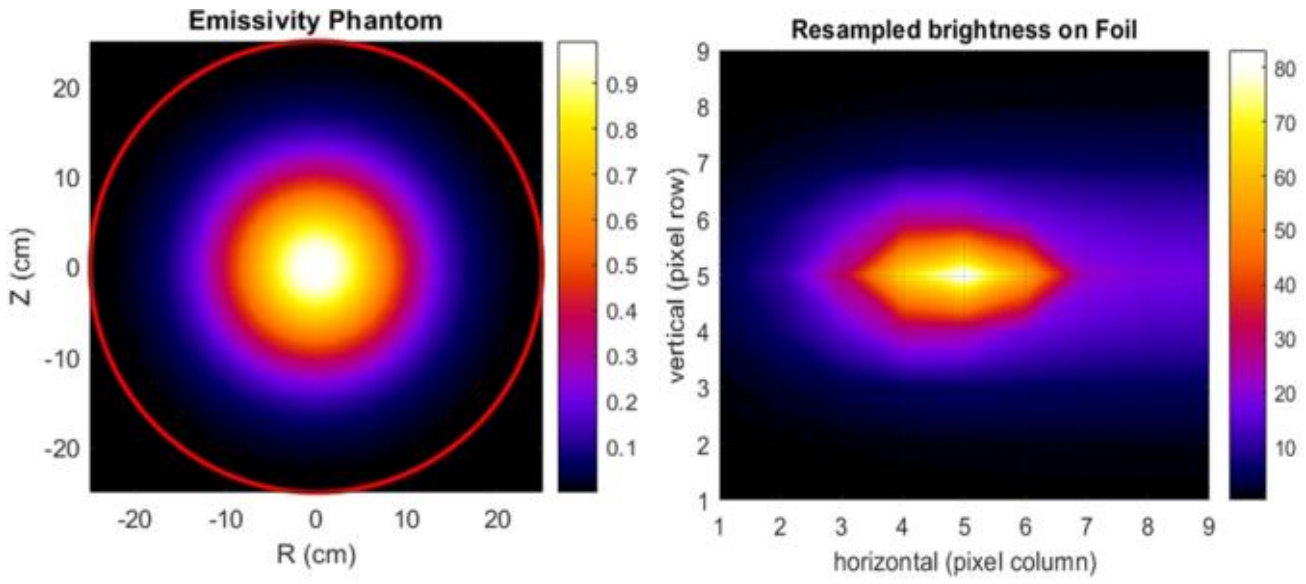

Figure 5. (a) Core Gaussian emissivity phantom in the R–Z plane. (b) Corresponding resampled brightness as viewed by the IRVB foil. Red circle denotes the plasma boundary.

### B. Hollow Gaussian Profile

The hollow Gaussian phantom models annular emission arising from hollow impurity density profiles or H-mode pedestal accumulation. Let $r \;=\; \sqrt{(R-R_0)^2+(Z-Z_0)^2}$:

$$\epsilon_2(R,Z) \;=\; \epsilon_0 \left(\frac{r}{r_0}\right)^2 exp\left[-\frac{(r-r_0)^2}{2{\sigma_{\mathrm{h}}}^2}\right] \tag{7}$$

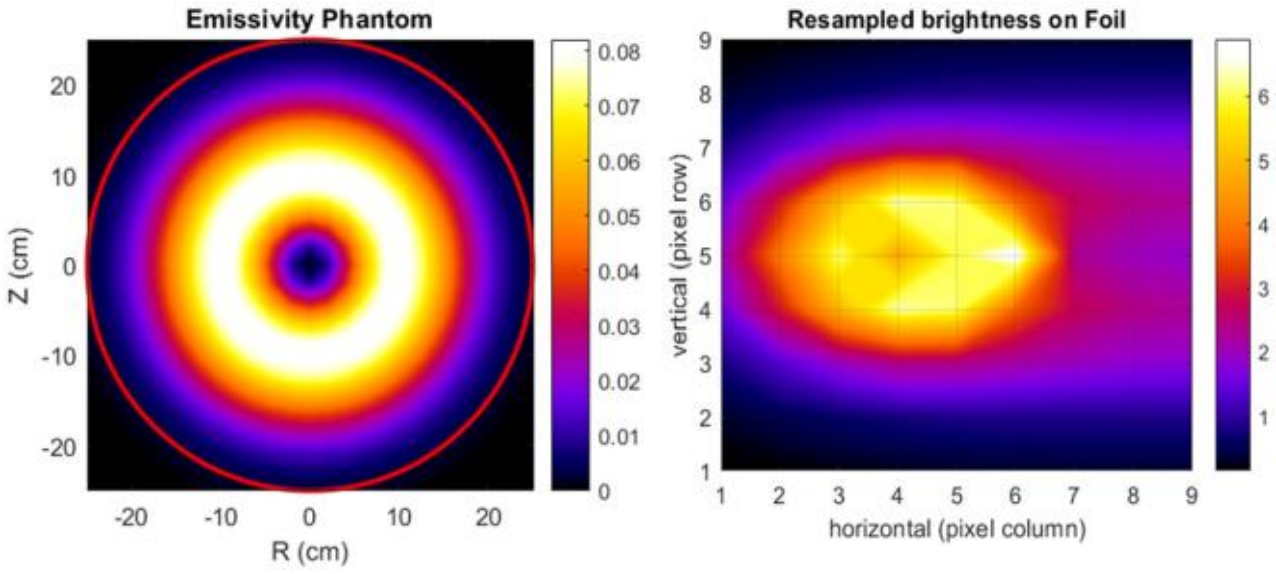

Figure 6. (a) Hollow Gaussian (annular ring) emissivity phantom. (b) Corresponding foil brightness distribution.

### C. Low-Field-Side Asymmetric Profile

This phantom model's outboard impurity accumulation driven by E×B drift imbalance as well as localized limiter plasma interaction:

$$\epsilon_3(R,Z) \;=\; \epsilon_3 exp\left[-\left(\frac{(R-R_0-\Delta R_1)^2}{2\sigma_R^2}+\frac{(Z-Z_0)^2}{2\sigma_Z^2}\right)\right]\Theta(R-R_0) \tag{8}$$

Where, $\Delta R_1 \;=\; 8\;cm$ is the outboard offset, $\sigma_R \;=\; 6\;cm$, $\sigma_Z \;=\; 10\;cm$, and $\Theta$ is the Heaviside step function enforcing one-sided emission.

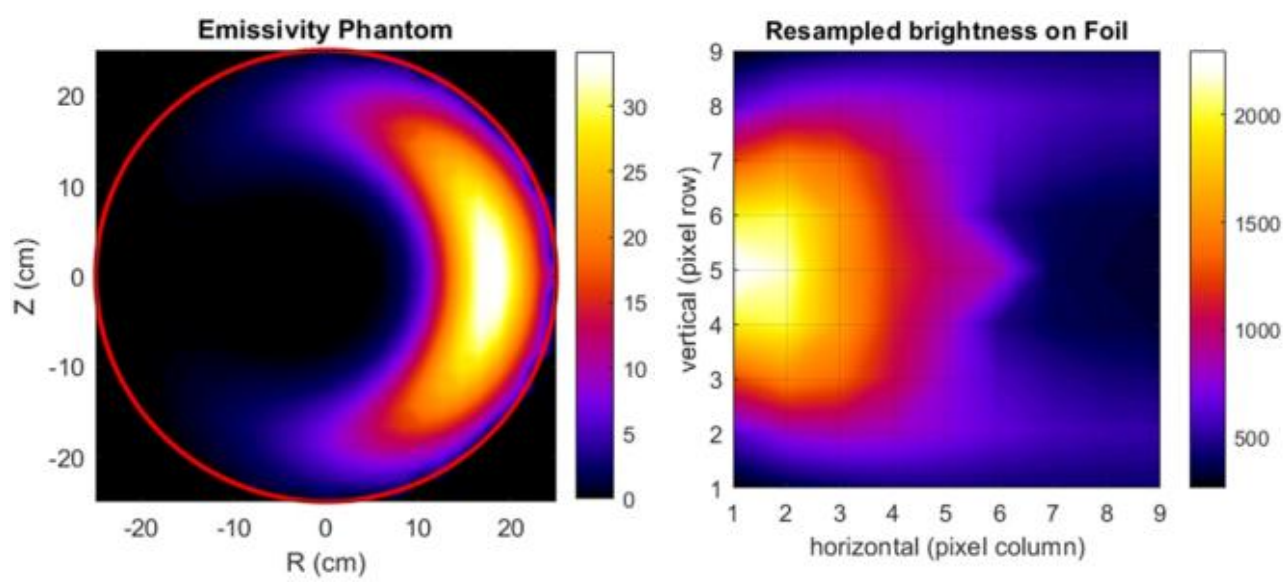

Figure 7. (a) Low-field-side asymmetric emissivity phantom. (b) Foil brightness showing the characteristic asymmetric distribution.

### *D. Double-Null Divertor Profile*

Two Gaussian blobs replicate emission at divertor strike points in a double-null configuration:

$$\epsilon_4(R,Z) \;=\; \sum_{k=1}^{2} \varepsilon_{\mathrm{k}}\, exp\left[-\frac{(R-R_{\mathrm{k}})^2+(Z-Z_{\mathrm{k}})^2}{2{\sigma_{\mathrm{k}}}^2}\right] \tag{9}$$

Upper blob: $(R_1, Z_1) = (-15, +18)\,cm$, $\sigma_1 = 4\,cm$, $\varepsilon_1 = 2.0$. Lower blob: $(R_2, Z_2) = (-15, -18)\,cm$, $\sigma_2 = 4\,cm$, $\varepsilon_2 = 1.8$.

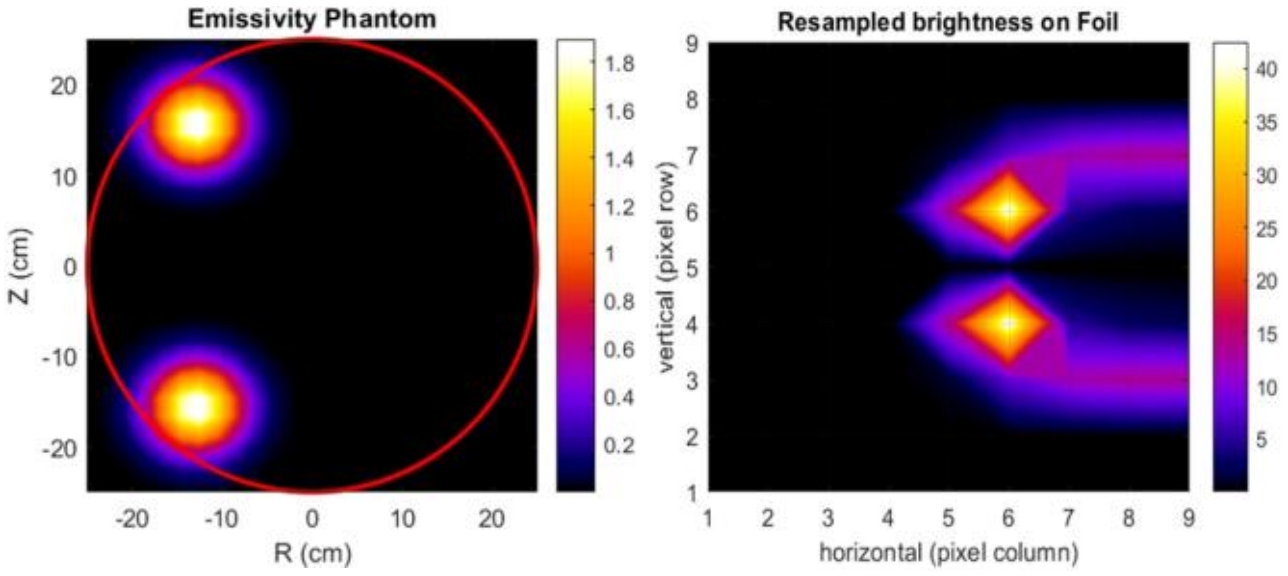

Figure 8. (a) Double-null divertor emissivity phantom (two Gaussian blobs). (b) Corresponding foil brightness showing two vertically separated peaks.

### *E. Impurity Emission Profile*

The electron temperature and density profiles corresponding to synthetic plasma profiles in a medium sized tokamak are prescribed as:

$$T_{\mathrm{e}}(R,Z) = T_{e0}(1-\rho^2)^{\alpha}\,, \tag{10}$$

$$n_e(R,Z) = n_{e0}(1-\rho^2)^{\beta} \tag{11}$$

Where, $\rho = \frac{r}{a_0}$ is the normalized minor radius, $a_0 = 25\,cm$ the plasma minor radius, $T_{e0} = 250\,eV$, and $n_{e0} = 2 \times 10^{13}\,cm^{-3}$. The parameters $\alpha$ and $\beta$ are parabola profile shaping factors for temperature profile and density profile respectively. In present case, values $\alpha = 1.75$ and $\beta = 1.0$ are assumed [33], [34], [35], [36], [37]. The volumetric emissivity is:

$$\epsilon_5(R,Z) \;=\; n_e^2(R,Z) \times \sum_z c_z\, \Lambda_z\big(T_{\mathrm{e}}(R,Z)\big) \tag{12}$$

where $c_z$ is the fractional concentration of impurity species z and $\Lambda_z(T_e)$ is the corresponding OpenADAS cooling rate coefficient [6]. Figure shows the cooling rates of different impurities expected in ADITYA-U tokamak, namely, Carbon (C), Oxygen (O), and Iron (Fe) and their variation with temperature ($T_e$) for a fixed density $n_{e0}$.

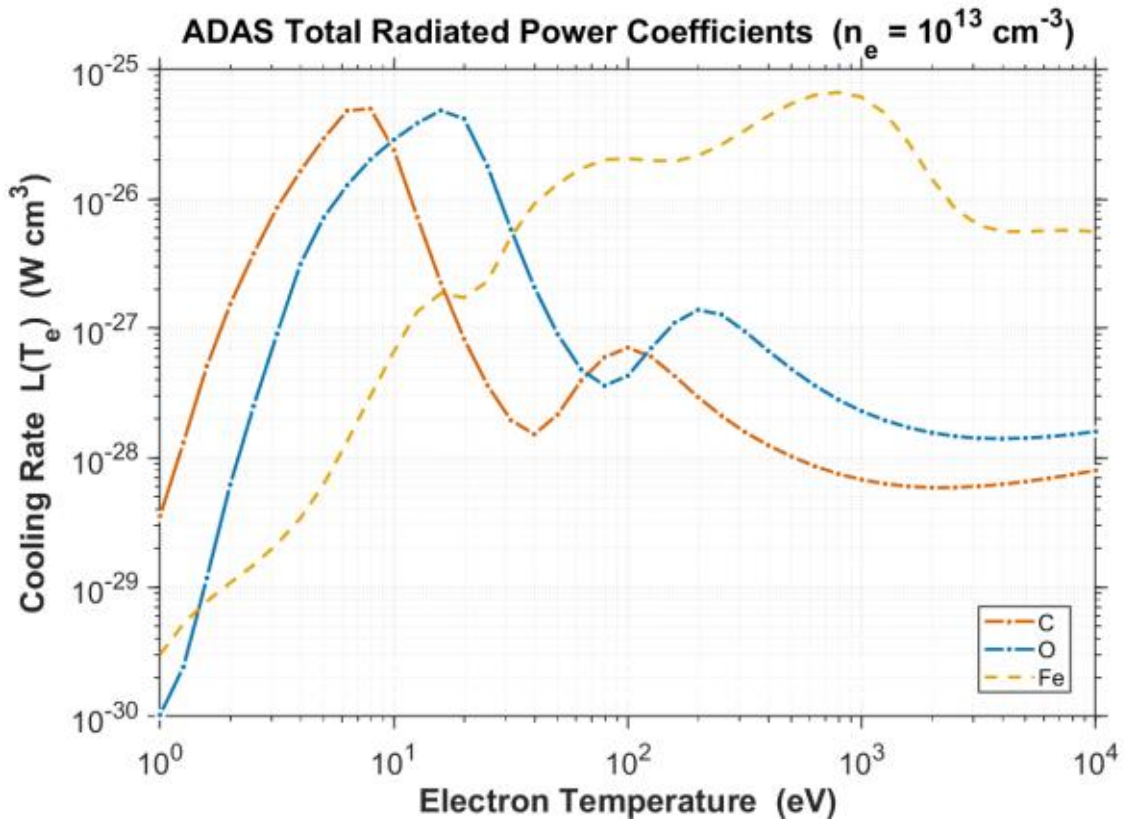

Figure 9. Cooling rates (in $W \cdot cm^3$) of C, O, and Fe[2] impurities plotted against electron temperature (in eV).

Expanding (12) for three species:

$$\epsilon_5 \;=\; n_e^2\,[c_C\Lambda_C \;+\; c_O\Lambda_O \;+\; c_{Fe}\Lambda_{Fe}\,] \tag{13}$$

Concentrations: $c_C = 3\%$, $c_O = 3\%$, $c_{Fe} = 0.05\%$ are assumed [39], [40], [41]. Despite its low concentration, Fe dominates the radiative loss owing to its substantially larger cooling rate coefficient at plasma core temperatures.

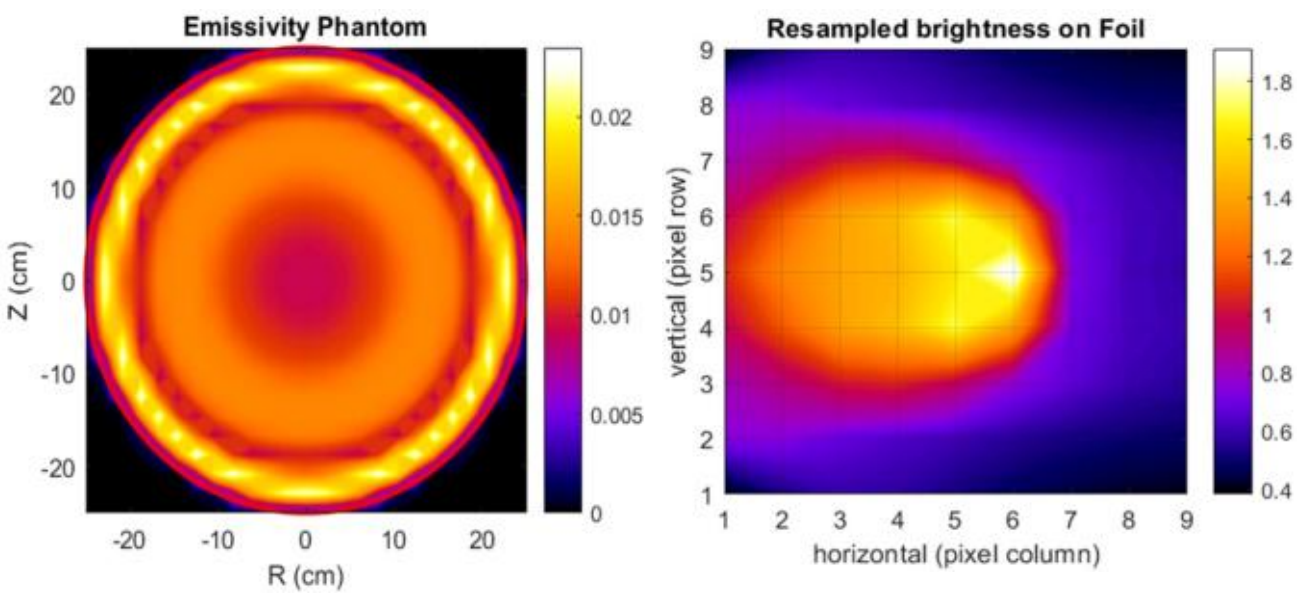

Fig. 5. (a) Physical impurity emission phantom ($3\%\,C + 3\%\,O + 0.05\%\,Fe$; $T_{\mathrm{e}} = 250\,eV, n_{\mathrm{e}} = 2 \times 10^{13} cm^{-3}$). (b) Corresponding foil brightness showing edge-dominated profile.

## IV. TOMOGRAPHIC RECONSTRUCTION METHODS

The four algorithms share the requirement of non-negativity ($\hat{\varepsilon} \geq 0$) but differ in how regularization is applied, whether a closed-form or iterative solution is obtained, and what statistical model is assumed for the measurement noise. Let $\Gamma = diag({\sigma_{\mathrm{i}}}^2)$ denote the noise covariance matrix, where $\sigma_{\mathrm{i}} = \alpha B_{\mathrm{i}}$ is the standard deviation of the $i^{th}$ measurement, $\alpha \in (0, 1]$ is the prescribed fractional noise level, and $B_{\mathrm{i}}$ is the $i^{th}$ line-integrated brightness measurement. Monte Carlo noise realisations are generated as ${B_{\mathrm{i}}}^{noisy} = B_{\mathrm{i}} + \sigma_{\mathrm{i}}\,\varepsilon_i$ where $\varepsilon_{\mathrm{i}} \sim \mathcal{N}(0,1)$ are standard normal variates.

### *A. First-Order Phillips-Tikhonov Regularization*

The first order PTR minimizes the penalized least-squares functional:

$$J_1(\varepsilon) \;=\; \left\|B - L_{proj}\cdot\varepsilon\right\|^2 \Gamma^{-1} \;+\; \lambda_1\|D_1\varepsilon\|^2 \tag{14}$$

where $D_1$ is the first-order finite-difference matrix approximating the spatial gradient in R and Z. Setting $\frac{\partial J_1}{\partial \varepsilon} = 0$ yields the closed-form normal equations:

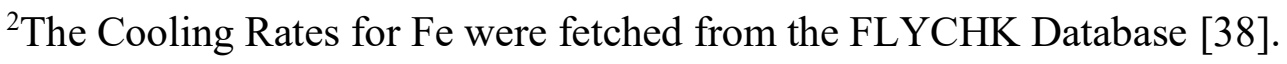

[2]The Cooling Rates for Fe were fetched from the FLYCHK Database [38].

$$\hat{\varepsilon}_{PTR_1} = \left(L_{proj}^T \Gamma^{-1} L_{proj} + \lambda_1 D_1^T D_1\right)^{-1} L_{proj}^T \Gamma^{-1} B \quad (15)$$

The regularization parameter $\lambda_1$ is selected via the L-curve criterion [7]. Non-negativity is enforced after the reconstruction. The solution is unique for any $\lambda_1 > 0$ and obtained in a single linear solve.

### *B. Second-Order Phillips-Tikhonov Regularization*

The second order PTR replaces the gradient penalty with a Laplacian penalty, enforcing higher-order smoothness:

$$J_2(\varepsilon) = \|B - L_{proj} \cdot \varepsilon\|^2 \Gamma^{-1} + \lambda_2 \|D_2 \varepsilon\|^2 \quad (16)$$

where $D_2$ approximates the discrete Laplacian $\nabla^2$ on the R–Z grid. The closed-form solution is:

$$\hat{\varepsilon}_{PTR_2} = \left(L_{proj}^T \Gamma^{-1} L_{proj} + \lambda^2 D_2^T D_2\right)^{-1} L_{proj}^T \Gamma^{-1} B \quad (17)$$

$D_2^T D_2$ is the discrete biharmonic operator, which penalizes curvature more strongly than PTR-1 (First-Order PTR) and yields piecewise-linear solutions more suited to moderate spatial variation.

### *C. Minimum Fisher Information*

The MFI method [8] introduces a spatially adaptive regularization weight inversely proportional to the current emissivity estimate, concentrating smoothness enforcement in low-emission regions:

$$J_{MFI}(\varepsilon) = \|B - L_{proj} \cdot \varepsilon\|^2 \Gamma^{-1} + \lambda_{MFI} \int \frac{|\nabla \varepsilon|^2}{\varepsilon} dV \quad (18)$$

, where $\lambda_{MFI}$ is the MFI regularization parameter, and $V$ is the volume of the toroidal chamber in which the emission is distributed. Discretizing the integral and defining $W(\varepsilon) = diag(1/\varepsilon_j)$:

$$J_{MFI}(\varepsilon) \approx \|B - L_{proj} \cdot \varepsilon\|^2 \Gamma^{-1} + \lambda_{MFI}\, \varepsilon^T D^T W(\varepsilon) D \varepsilon \quad (19)$$

Because W depends on ε, minimization is nonlinear and solved by iteratively reweighted least squares (IRLS). At iteration n, $W^n = diag(1/\varepsilon_n_j)$ is fixed, giving the update:

$$\varepsilon^{n+1} = \left(L_{proj}^T \Gamma^{-1} L_{proj} + \lambda_{MFI}\, D^T W^n D\right)^{-1} L_{proj}^T \Gamma^{-1} B \quad (20)$$

Convergence is declared when the relative update $\frac{\|\varepsilon n+1 - \varepsilon n\|}{\|\varepsilon n\|} < 10^{-4}$ or after 5 IRLS iterations. A floor $\varepsilon_{min} = 10^{-6}$ prevents division by zero in W.

### *D. Maximum-Likelihood Expectation-Maximization*

MLEM maximizes the Poisson log-likelihood of the measurements, inherently enforcing non-negativity through multiplicative updates [9]:

$$\varepsilon_j^{n+1} = \left(\frac{\varepsilon_i^n}{\Sigma_i L_{ij}}\right) \times \Sigma_i L_{ij} \left[\frac{B_i}{\left(L_{proj} \cdot \varepsilon^n\right)_i}\right] \quad (21)$$

where $\Sigma_i L_{ij}$ is the geometric sensitivity of voxel j and $\left(L_{proj} \cdot \varepsilon^n\right)_i = \Sigma_i L_{ij} \varepsilon_i^n$ is the forward-projected brightness at iteration *n*. Starting from a uniform estimate $\varepsilon_i^0 = 1$, the algorithm is stopped after 200 iterations or at relative change < $10^{-5}$. Early stopping provides implicit regularization; the iteration count controls the resolution-noise trade-off.

## V. RESULTS AND DISCUSSION

Figure 6 presents the emissivity reconstructed and Table II - Table III shows that MLEM achieves the lowest average computation time (0.36 s), owing to the simplicity of its element-wise multiplicative update despite requiring up to 200 iterations per reconstruction. PTR-2 follows at 0.62 s, and PTR-1 at 0.58 s both requiring a single sparse linear solve. MFI is the most computationally demanding method (3.02 s average) due to its IRLS inner loop of 10–20 iterations per reconstruction. However, MFI computation time remains within the range acceptable for inter-shot or offline analysis on present-day tokamaks. For real-time applications with cycle times on the order of 1 s, PTR-2 represents the optimal choice: closed-form computation below 0.7 s with second-best reconstruction fidelity across all phantoms.

TABLE II presents a thorough quantitative performance benchmarks for all four methods across all five phantoms under 5% Monte Carlo noise. Regularization parameters for PTR-1, PTR-2, and MFI are selected via the L-curve criterion[42] whereas MLEM uses the stopping rule defined in Section IV-D.

Table II

RELATIVE RECONSTRUCTION ERROR (%) AND STRUCTURAL CORRELATION COEFFICIENT (ρ) FOR ALL METHOD - PHANTOM COMBINATIONS.

| Method | Gaussian | Hollow | Banana-LFS | Double-Null | Impurity | Avg. |
|---|---|---|---|---|---|---|
| | | | ***Relative Error (%)*** | | | |
| PTR-1 | 8.53 | 19.62 | 12.84 | 53.65 | 34.60 | 25.85 |
| PTR-2 | **2.75** | 9.82 | **6.93** | 43.05 | **31.05** | 18.72 |
| MFI | 3.59 | **9.79** | 13.69 | **16.95** | 31.51 | **15.10** |
| MLEM | 25.11 | 38.21 | 16.92 | 31.06 | 44.18 | 31.09 |
| | | | ***Correlation Coefficient (ρ)*** | | | |
| PTR-1 | 0.996 | 0.957 | 0.988 | 0.845 | 0.425 | 0.842 |
| PTR-2 | **0.999** | **0.990** | **0.997** | 0.912 | **0.580** | 0.896 |
| MFI | **0.999** | **0.990** | 0.988 | **0.992** | **0.580** | **0.910** |
| MLEM | 0.946 | 0.805 | 0.981 | 0.953 | 0.087 | 0.755 |

*Note: Bold values indicate the best (lowest error / highest correlation) per phantom column. Phantom labels: Gaussian = core Gaussian; Hollow = hollow Gaussian; Banana-LFS = low-field-side asymmetric; Double-Null = DND divertor; Impurity = physical impurity emission profile.*

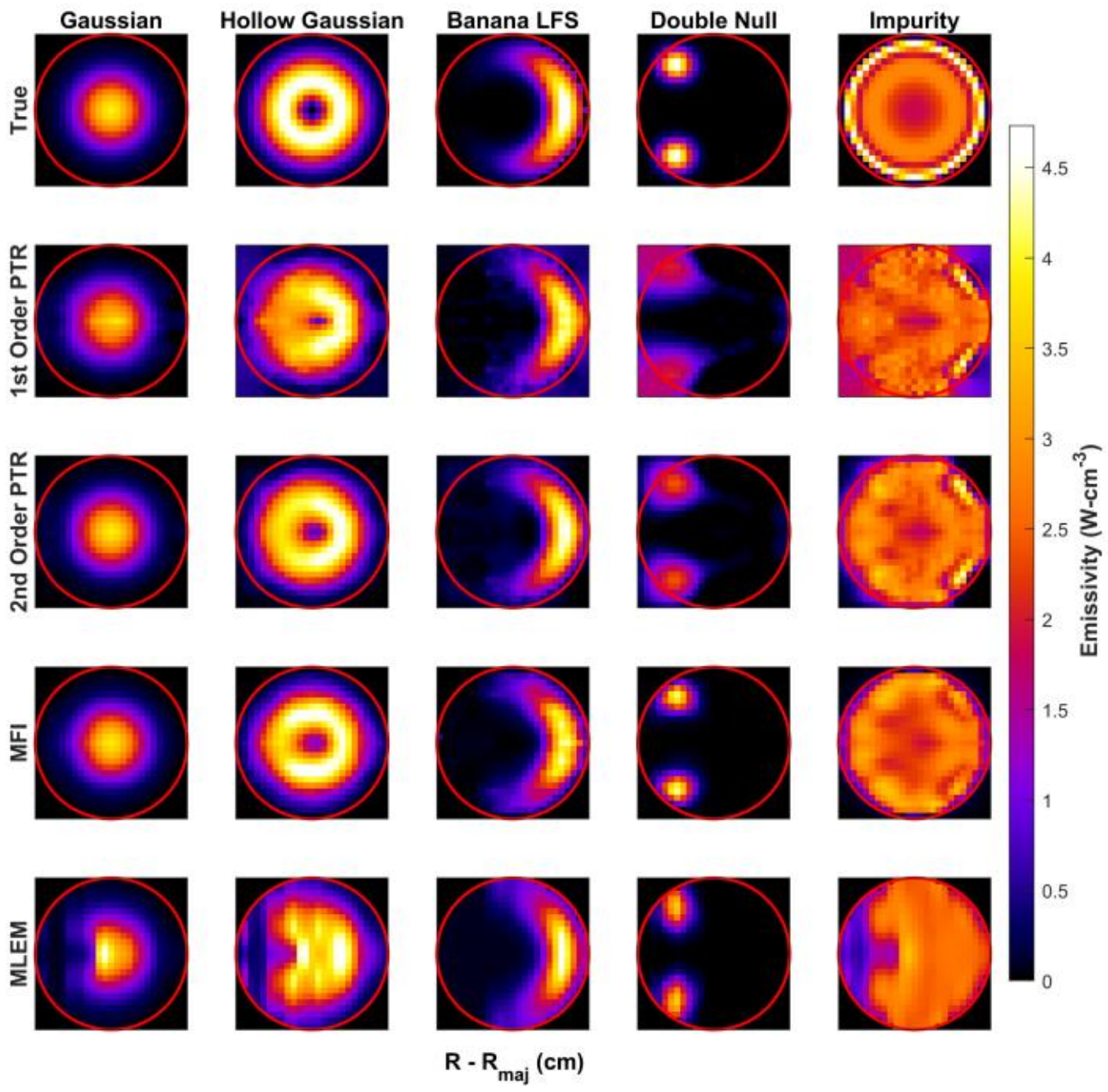

Figure 6. Tomographic reconstructions for all phantoms using MFI, PTR-1, PTR-2, and MLEM algorithms.

### *A. Relative Error and Structural Correlation*

As reported in Table II, MFI achieves the lowest average relative reconstruction error of 15.10% across all five phantoms, with PTR-2 as the closest competitor at 18.72%. The performance advantage of MFI is most pronounced for the double-null divertor phantom (16.95% vs. 43.05% for PTR-2), confirming that its spatially adaptive $1/\varepsilon$ regularization weight provides critical benefit for compact, spatially separated emission structures. PTR-1 performs well only for the core Gaussian phantom (8.53%) but degrades severely for the double-null divertor (53.65%), owing to the fixed gradient penalty smearing the compact blobs. MLEM incurs the highest average error (31.09%) and, critically, exhibits an extremely low correlation coefficient of 0.87 for the impurity phantom, with failure to recover the edge-dominated emission in this case which can be a consequence of noise amplification at the plasma boundary in the multiplicative update step.

Regarding structural correlation, MFI achieves the highest average (0.910), marginally above PTR-2 (0.896). Both MFI and PTR-2 attain $\rho = 0.999$ for the core Gaussian phantom. The most discriminating phantom is the impurity emission profile, for which PTR-2 and MFI both achieve $\rho = 0.580$ while MLEM collapses to 0.087 - a difference of more than one standard deviation in reconstruction quality that carries direct diagnostic implications for radiation boundary monitoring.

### *B. Computational Performance*

Table III shows that MLEM achieves the lowest average computation time (0.36 s), owing to the simplicity of its element-wise multiplicative update despite requiring up to 200 iterations per reconstruction. PTR-2 follows at 0.62 s, and PTR-1 at 0.58 s both requiring a single sparse linear solve. MFI is the most computationally demanding method (3.02 s average) due to its IRLS inner loop of 10–20 iterations per reconstruction. However, MFI computation time remains within the range acceptable for inter-shot or offline analysis on present-day tokamaks. For real-time applications with cycle times on the order of 1 s, PTR-2 represents the optimal choice: closed-form computation below 0.7 s with second-best reconstruction fidelity across all phantoms.

TABLE III

RECONSTRUCTION COMPUTATION TIME (SECONDS) FOR ALL METHOD-PHANTOM COMBINATIONS.

| Method | Gaussian | Hollow | Banana-LFS | Double-Null | Impurity | Avg. |
|---|---|---|---|---|---|---|
| 1st PTR | 0.61 | 0.62 | 0.54 | 0.54 | 0.61 | 0.58 |
| 2nd PTR | 0.55 | 0.62 | 0.69 | 0.63 | 0.61 | 0.62 |
| MFI | 2.99 | 3.18 | 3.06 | 2.85 | 3.01 | 3.02 |
| MLEM | **0.49** | **0.32** | **0.32** | **0.31** | **0.34** | **0.36** |

*Note: Bold values indicate the fastest computation per phantom column. MLEM achieves the lowest average computation time (0.36 s) owing to its simple element-wise multiplicative update, despite requiring up to 200 iterations. MFI is the most computationally expensive method owing to IRLS inner iterations.*

## VI. SENSITIVITY ANALYSIS

The reconstruction accuracy of IRVB tomography is inherently coupled to two discretization parameters: the number of bolometer channels N×N, which determines the number of independent measurements available to the inversion, and the resolution of the R-Z reconstruction grid. In the baseline configuration studied in Section V, these are fixed at 9×9 and 25×25, respectively, chosen to represent the ADITYA IRVB setup[7]. The following two subsections systematically examine how reconstruction accuracy and computation time respond to independent variation in each of these parameters across all five phantoms and all four algorithms.

### *A. Effect of Bolometer Channel Count on Reconstruction Accuracy*

The reconstruction grid is fixed at 25×25 and the camera resolution is swept across 17 configurations from N = 4 to N = 20 (i.e., N×N channels), under 5% Monte Carlo noise. Figure 7 presents the average $\varepsilon_{rel}$ across all five phantoms as a function of channel count, and Figure 8 shows the per-phantom breakdown for each method.

All regularization-based methods (PTR-1, PTR-2, MFI) exhibit a broadly monotonic decrease in $\varepsilon_{rel}$ with increasing channel count, consistent with the improved conditioning of the geometry matrix $L_{proj}$ as the measurement-to-unknown ratio increases. At the sparse end (N = 4), errors are large for all methods: PTR-1 averages 45.5%, PTR-2 averages 40.3%, MFI averages 34.5%, and MLEM averages 50.5%. By the baseline configuration N = 9, these reduce to approximately 26%, 20%, 16.5%, and 33.5% respectively, consistent with the main results of Section V. At highest channel count (N = 20), PTR-2 achieves the lowest average error at approximately 8%, followed by PTR-1 at 11.5%. Notably, MFI's average error rises to approximately 27% at N = 20, driven entirely by the impurity emission phantom for which its adaptive $1/\varepsilon$ weight amplifies noise at the plasma boundary as the measurement density increases which is a known sensitivity of Fisher-information regularization in high-SNR, edge-dominated regimes [43]. For the remaining four phantoms, MFI continues to outperform PTR-1 at all channel counts and maintains near-parity with PTR-2.

The per-phantom breakdown in Figure 8 reveals several radiative-feature specific trends. For the core Gaussian and hollow Gaussian phantoms, all methods except MLEM converge below 15% error by N = 9 to 12. For the banana-LFS and double-null divertor phantoms which are the most geometrically demanding cases, the benefit of increasing channel count is most pronounced: PTR-2's error on the banana-LFS drops from 27% at N = 4 to 5% at N = 20, and MFI's error on the double-null divertor improves from 57% at N = 4 to 9% at N = 20. MLEM improves overall but remains the highest-error method at all channel counts, with no configuration achieving average $\varepsilon_{rel}$ below 27%. These results confirm that the principal bottleneck for MLEM is not measurement sparsity but rather the noise-amplifying character of its multiplicative update, which persists regardless of channel count.

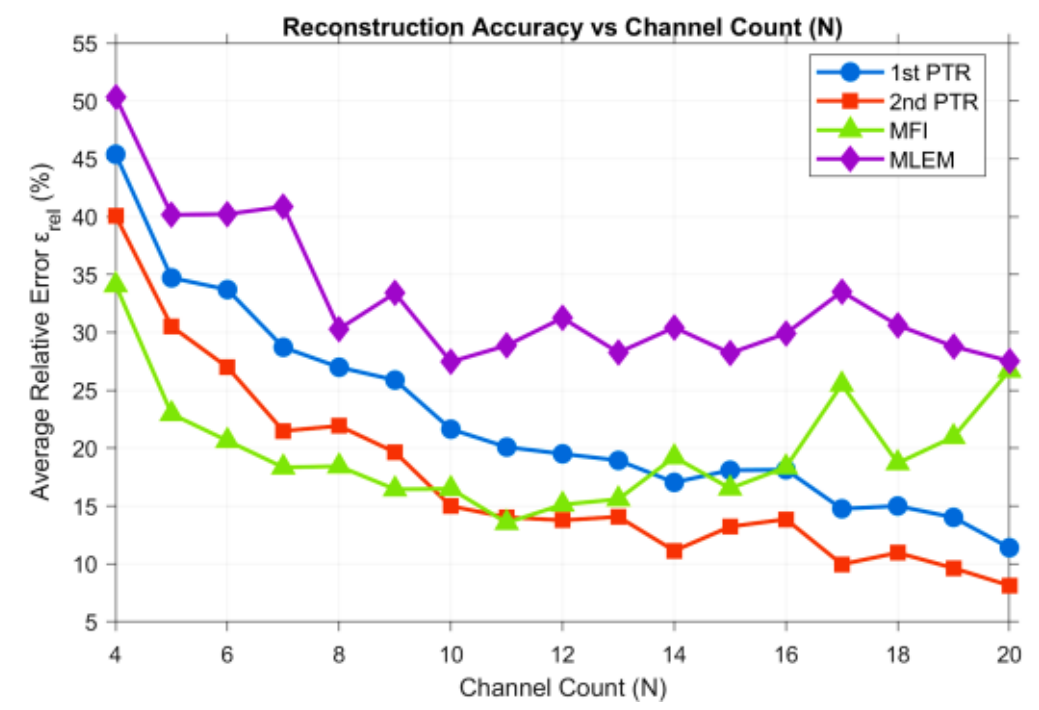

Figure 7. Average relative reconstruction error $\varepsilon_{rel}$ (%) averaged across all five phantoms as a function of bolometer camera channel count N (N×N pixels), with the R–Z reconstruction grid fixed at 25×25.

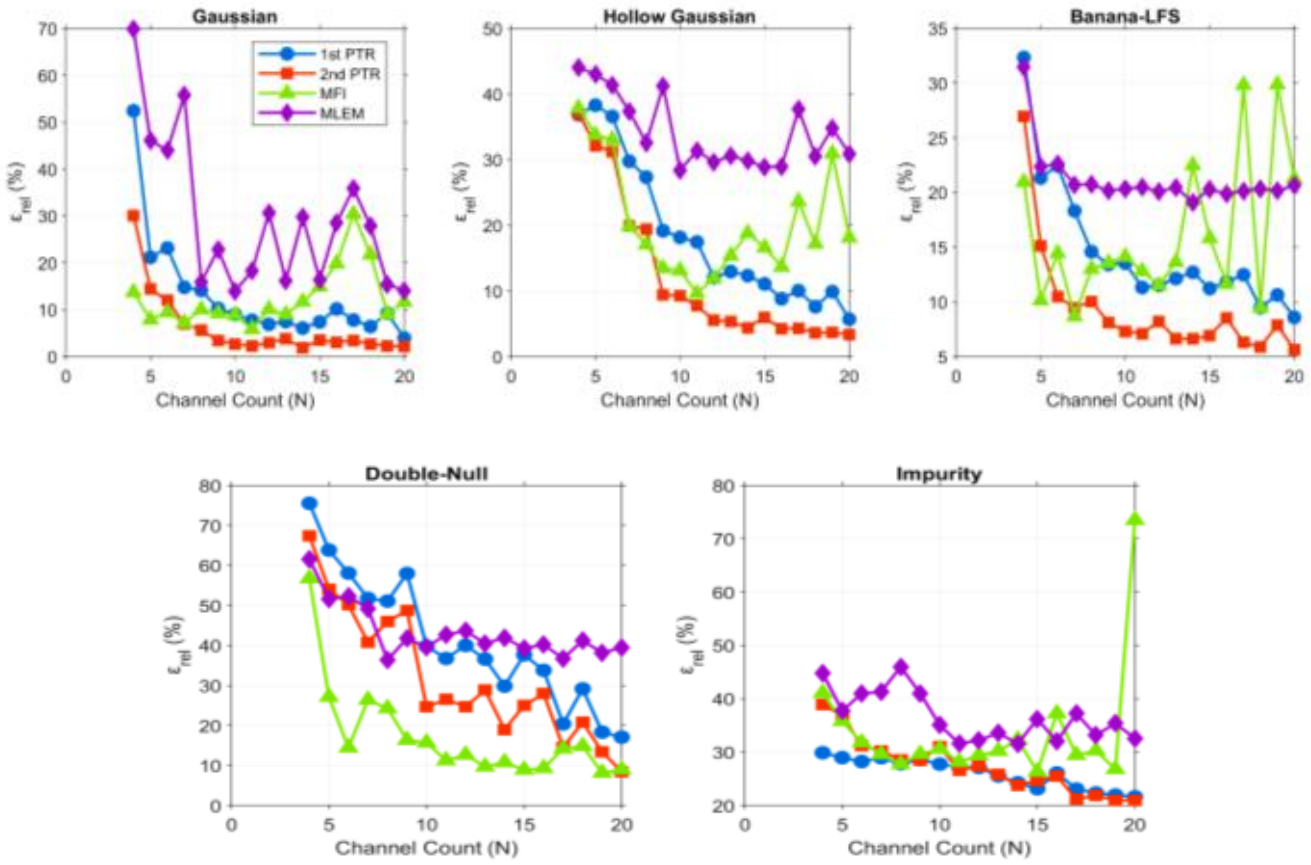

Figure 8. Per-phantom relative reconstruction error $\varepsilon_{rel}$ (%) as a function of channel count N for all five phantoms: (a) Gaussian, (b) Hollow Gaussian, (c) Banana-LFS, (d) Double-Null, (e) Impurity.

Computation time as a function of channel count is shown in Figure 9. PTR-1, PTR-2, and MLEM all remain below 0.75 s across the entire study, as their per-reconstruction cost scales primarily with the fixed grid dimension (625 unknowns) rather than the number of measurements. MFI rises from approximately 2.0 s at N = 4 to 3.8 s at N = 20, reflecting the increased cost of forming and solving the $N^2 \times P$ IRLS system as the measurement vector grows. Nevertheless, all methods remain well within the inter-shot analysis time budget of practical tokamak operation at all tested configurations.

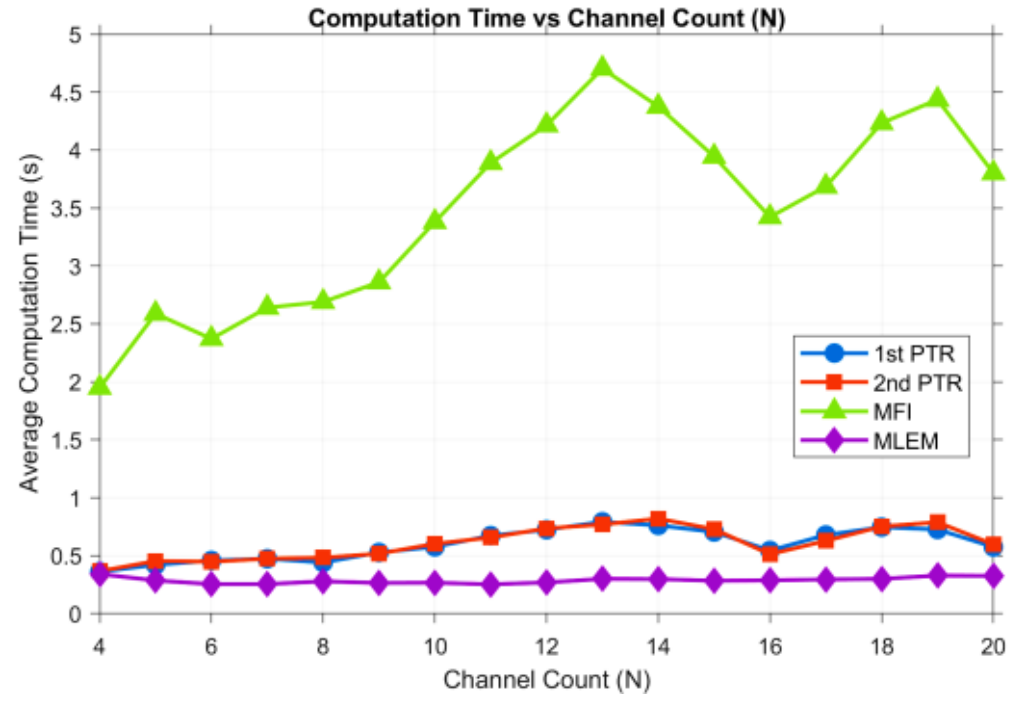

Figure 9. Average computation time (s) as a function of channel count N, with reconstruction grid fixed at 25×25.

### *B. Effect of R-Z Grid Resolution on Reconstruction Accuracy*

The camera is fixed at the baseline 9×9 configuration and the R-Z reconstruction grid is swept across 17 resolutions from N = 11 to N = 41 with a step of 2 (i.e., $N_R \times N_Z$ voxels in the poloidal plane). The choice of odd-numbered voxel grid was chosen entirely to include the contribution from the central row of bolometer pixel grid. As the grid becomes finer, the number of unknowns grows as $N^2$ while the number of measurements remains fixed at 81, so the inverse problem becomes progressively more underdetermined beyond a certain resolution threshold. Figure 10 shows the average $\varepsilon_{rel}$ across all phantoms as a function of grid size, and Figure 11 shows the per-phantom average errors.

At the coarsest grid (N = 11, i.e., 121 unknowns), the inverse problem is over-determined relative to the 81 available measurements, and all methods produce relatively accurate results. PTR-2 achieves the lowest average error of approximately 7.5%, followed by PTR-1 at 12% and MFI at 13%, while MLEM remains elevated at 38%. As the grid is refined to N = 15, the ratio of unknowns to measurements increases, and the errors of PTR-1 and PTR-2 rise substantially, PTR-1 reaches 25% at N = 15, and PTR-2 climbs to 23%, and MLEM drops to 35.6%. MFI, however, remains comparatively stable in the range 15-17% from N = 17 to N = 41 owing to its adaptive weight matrix, which maintains effective regularization strength despite the growing underdetermination.

Beyond N = 25, average errors stabilize for all methods: PTR-2 fluctuates between 16-19%, PTR-1 between 25-28%, MFI between 15-17%, and MLEM between 33-35%. No meaningful accuracy improvement is obtained by increasing the grid beyond 25×25 for the 9×9 camera. This confirms that the baseline choice of a 25×25 grid lies at or near the accuracy saturation point: it provides sufficient spatial resolution to resolve the phantoms studied while remaining well-conditioned relative to the available 81 measurements. Grids coarser than 20×20 provide a better measurement-to-unknown ratio but sacrifice the spatial resolution needed to represent compact emission features such as the double-null divertor blobs.

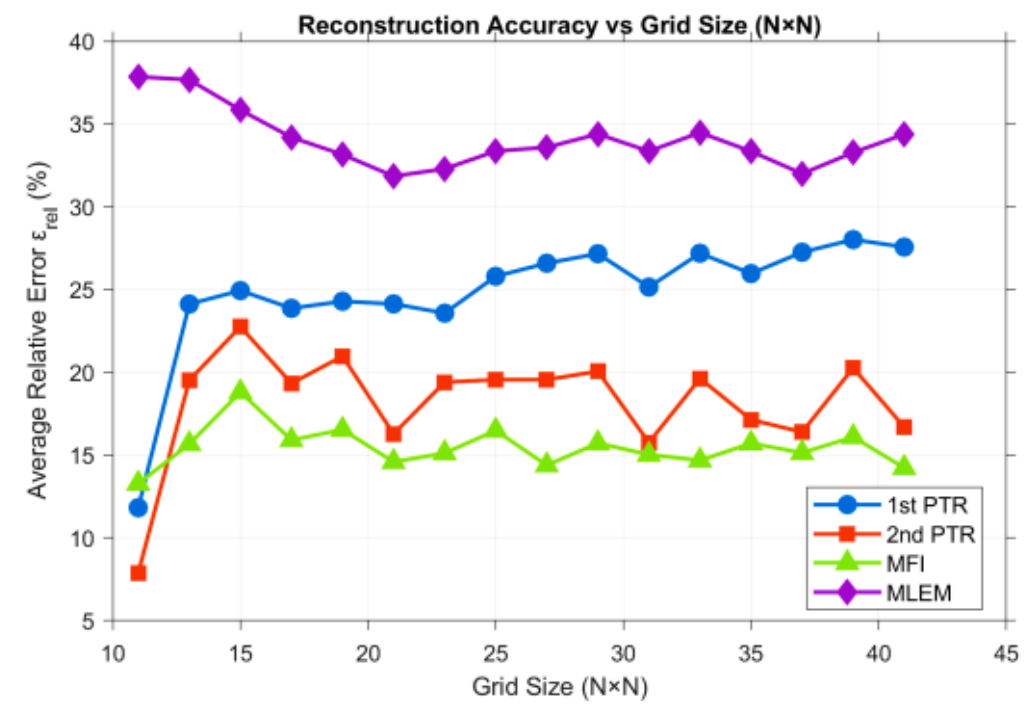

Figure 10. Average relative reconstruction error $\varepsilon_{rel}$ (%) averaged across all five phantoms as a function of R–Z grid size N×N, with camera fixed at 9×9 channels.

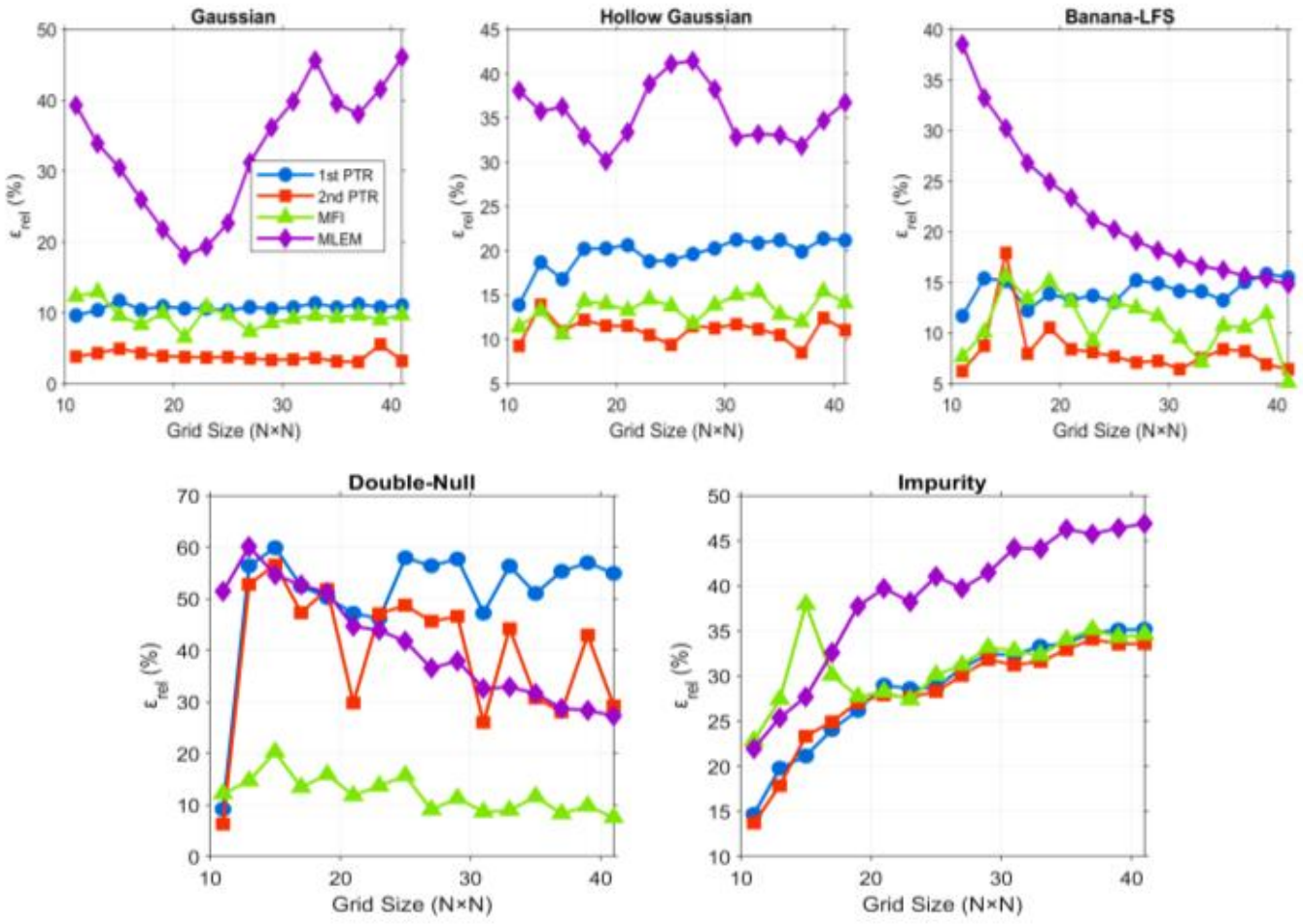

Figure 11. Per-phantom relative reconstruction error $\varepsilon_{rel}$ (%) as a function of R–Z grid size N×N for all five phantoms: (a) Gaussian, (b) Hollow Gaussian, (c) Banana-LFS, (d) Double-Null, (e) Impurity.

The per-phantom results in Figure 11 reveal that the impurity and double-null divertor phantoms are the most sensitive to grid resolution choice. For the double-null divertor, MFI achieves errors below 21% at all grid sizes examined, while PTR-1 ranges from 10% at N = 11 to 59% at N = 15 before partially recovering, a non-monotonic behaviour caused by the interaction between grid discretization and the ability to represent two compact blobs located near the plasma boundary. For the impurity phantom, the errors of all methods increase gradually with grid size, which is consistent with the expectation that edge-peaked profiles are more difficult to reconstruct as the boundary voxels become smaller and the number of boundary-adjacent grid points grows, producing increased sensitivity to boundary regularization conditions.

The computation time scaling with grid size (Figure 12) is the consequential result of Study B. PTR-1 and PTR-2, which require a linear solve of dimension $N^2 \times N^2$, scale roughly as $O(N^6)$ in the dense-solver limit, rising from approximately 0.1 s at N = 11 to 3.4 s at N = 41. MLEM remains nearly constant at 0.1-0.4 s across all grid sizes, as its per-iteration cost scales only linearly with the number of voxels. MFI, however, exhibits the steepest scaling: its average computation time rises from 0.5 s at N = 11 to 17.7 s at N = 41, a 35-fold increase over roughly a factor of four in grid linear dimension, consistent with the $O(N^6)$ cost of the repeated IRLS solves combined with the dense Fisher weight matrix computation. This scaling makes MFI impractical at grid resolutions above approximately 30×30 for the current IRLS implementation without further algorithmic optimisation such as sparse approximations of the Fisher weight matrix or conjugate-gradient-based iterative solvers.

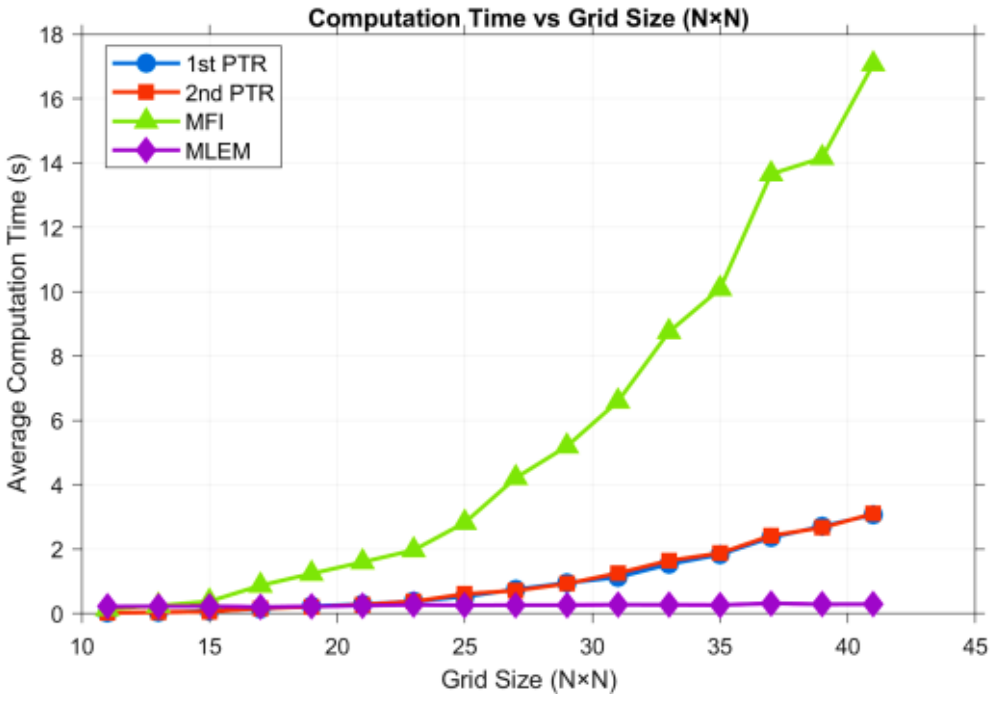

Figure 12. Average computation time (s) as a function of R–Z grid size N×N, with camera fixed at 9×9 channels.

## VII. CONCLUSION

A systematic forward-modelling and inverse-reconstruction framework has been presented and benchmarked for IRVB tomography across five synthetic emissivity phantoms and four reconstruction algorithms executed on a Dell Precision 5820 Tower workstation equipped with an Intel Xeon W-2245 processor (3.90 GHz, 8 cores, 16 logical processors) and 32 GB RAM running Microsoft Windows 11 Pro for Workstations. MFI consistently achieves the lowest reconstruction errors (2.75–16.95%), with its spatially adaptive 1/ε regularization providing the greatest benefit for compact and asymmetric emission structures such as the double-null divertor configuration. For offline high-fidelity reconstruction where computation time is unconstrained, MFI is therefore suitable for IRVB tomography. PTR-2 offers the best computational efficiency (0.62 s) with reconstruction errors within 3–5 percentage points of MFI, making it the preferred algorithm for near-real-time inter-shot analysis with cycle times below 1 s. PTR-1, while adequate for smooth core-peaked profiles, degrades markedly for non-Gaussian emission including divertor and asymmetric impurity distributions owing to its fixed gradient penalty, and is not recommended for diagnostically demanding configurations. MLEM, despite achieving the fastest average computation time (0.36 s) and inherently enforcing non-negativity, should be reserved for applications in which Poisson measurement statistics are physically appropriate, as it exhibits poor reconstruction fidelity for edge-dominated emission profiles. Future work will incorporate the effects of 3D modelling of VOS from IRVB channels on the tomographic reconstruction, as well as comparison of algebraic techniques to trained ML models and apply the framework to experimental IRVB data from limiter and diverted plasma configurations.